abstractabstract

# Network analysis of geomagnetic substorms using the SuperMAG database of ground-based magnetometer stations

J. Dods[1], S. C. Chapman[1,2], and J. W. Gjerloev[3,4]

[1]Centre for Fusion, Space and Astrophysics, Department of Physics, University of Warwick, Coventry, UK, [2]Department of Mathematics and Statistics, UIT, Tromsø, Norway, [3]Department of Physics and Technology, University of Bergen, Bergen, Norway, [4]The Johns Hopkins University Applied Physics Laboratory, Laurel, Maryland, USA

**Abstract** The overall morphology and dynamics of magnetospheric substorms is well established in terms of the observed qualitative auroral features seen in ground-based magnetometers. This paper focuses on the quantitative characterization of substorm dynamics captured by ground-based magnetometer stations. We present the first analysis of substorms using dynamical networks obtained from the full available set of ground-based magnetometer observations in the Northern Hemisphere. The stations are connected in the network when the correlation between the vector magnetometer time series from pairs of stations within a running time window exceeds a threshold. Dimensionless parameters can then be obtained that characterize the network and by extension, the spatiotemporal dynamics of the substorm under observation. We analyze four isolated substorm test cases as well as a steady magnetic convection (SMC) event and a day in which no substorms occur. These test case substorms are found to give a consistent characteristic network response at onset in terms of their spatial correlation. Such responses are differentiable from responses to the SMC event and nonsubstorm times. We present a method to optimize network parametrization with respect to the different individual station responses, the spatial inhomogeneity of stations in the Northern Hemisphere, and the choice of correlation window sizes. Our results suggest that dynamical network analysis has potential to quantitatively categorize substorms.

## 1. Introduction

Substorms are an extensively studied phenomena in geophysics, and while there is an overall established substorm cycle [*McPherron et al.*, 1973], there is considerable variation in the specific detailed sequence of events [*Akasofu*, 2004; *Meng and Liou*, 2004]. The ability to quantify substorm dynamics in an automated manner would be a valuable tool to determine what initial conditions, in terms of the internal state of the magnetosphere-ionosphere system and energy loading by the solar wind, produce a given detailed response. Attempts have been made at a classification of substorms based on images of the aurora [*Syrjasuo et al.*, 2007], where training algorithms were used to identify a wide range of arc shapes that can be present during a substorm. *AE* indices have also been used to identify substorm behavior [*Gjerloev et al.*, 2004], although such descriptions are limited by the scalar and spatially aggregating nature of the *AE* indices.

Ground-based magnetometer stations detect the variation in the local magnetic field resulting from time-dependent current systems in the ionosphere and serve as a proxy for dynamics occurring in the magnetosphere. There are typically ∼100 magnetometer stations available to observe any given substorm. The question is whether an algorithmic methodology can be developed to quantitatively characterize a substorm signature from these ∼100 time series in an automated manner. SuperMAG is a database that collates and processes all available ground-based magnetometer vector time series into a standardized baselined format at 1 min cadence [*Gjerloev*, 2012]. This provides an excellent starting point for studies attempting to characterize collective information from these stations. Here we investigate whether canonical correlation between the vector time series of pairs of magnetometer stations can be used to construct a network that can characterize substorms. Correlation between stations has been examined previously, although only using a few contraposed station pairs [*Jackel et al.*, 2001]. If the correlation between all ∼100 stations can be quantified in a robust and readily accessible manner, then this would provide a tool for substorm identification and classification.

The use of complex networks [*Boccaletti et al.*, 2006] has expanded from its birth in the social sciences [*Milgram*, 1967] through uses in biological [*Nicol et al.*, 2012] and engineering systems [*Sivrikaya and Yener*, 2004] to geophysical systems [*Radebach et al.*, 2013; *Malik et al.*, 2012; *Donges et al.*, 2009]. Recent work on spatially embedded networks [*Heitzig et al.*, 2012] has facilitated the application of complex networks to physical systems where the observations are spatially distributed. A dynamical network is defined by instances of time-varying connections that exist between the available pairs of points (here the magnetometer stations). In a substorm context the connections are identified using similarities between the vector time series of pairs of stations. Importantly, pairs of stations are either connected or not connected so that at its core, network methodology consists of quantifying when the cross correlation between the time series of a given pair of stations exceeds a threshold. A central problem in applying this methodology to real-world systems is identifying the appropriate threshold. This depends on the individual station responses which are influenced by multiple factors such as ground conductivity, oceanic currents, and instrument response. We will present new methodology to identify the threshold needed to account for such effects. Once the thresholds are determined, a series of networks can be constructed for a given substorm and parameters formed to describe the network topology. Questions then arise: do these dimensionless network parameters robustly identify the substorms? If so, then could this provide a new framework to quantitatively identify and characterize geomagnetic substorm activity? Is there a statistical network that describes the statistical substorm?

In this paper we establish a new methodology for dynamical network analysis in the magnetospheric substorm context and identify several dimensionless network parameters that capture key aspects of the dynamics of the substorm. A brief overview of the paper is as follows: In section 2 we introduce the data set used here, how connections in our network of stations were determined, and how the dimensionless network parameters that are used to describe the substorms are formed. In section 3 we apply this methodology to the four test case substorms as well as a steady magnetic convection (SMC) event and a day in which no substorms occur. The methodology is outlined in the body of the paper and further details are provided in the supporting information (SI) section.

## 2. The Data Sets Used in This Study

We used vector magnetometer time series data at 1 min cadence from the SuperMAG database. The SuperMAG database is an amalgamation of stations from different magnetometer station groups. All magnetometer data have been preprocessed in an identical way to remove long-term (> 1 day) trends [*Gjerloev*, 2012]. The vector time series are in local magnetic coordinates [*Gjerloev*, 2012]. Care must be taken due to variations in resolution, dynamical range, and local ground conductivity [*Tanskanen et al.*, 2001] between magnetometers from different groups, as we discuss briefly in section 2.1 (with more detail in SI Text S1). The nonlocal time-dependent contributions from ground conductivity cannot be removed from the magnetometer time series. However, the contributions are accounted for here in terms of their average effect on correlation between stations through the choice of station-specific thresholds. All active sites in the Northern Hemisphere between magnetic latitude (MLAT) 50° to 90° are used in the analysis. The distribution of available stations below 50° MLAT becomes much more inhomogeneous, such inhomogeneities can distort network parameters.

Four substorms are investigated and were selected according to the criteria outlined in *Gjerloev and Hoffman* [2014]: temporal isolation, the substorms do not occur during a magnetic storm (|DST| < 30 nT), and they are of the classic bulge type. They occur in the years 1997 and 1998 November–February. Winter months were chosen to limit sunlight on the dayside. The events are also selected such that there is a good distribution of stations in the nightside at the onset of the substorm. Onset times for the substorms were determined to 1 min precision using the Polar satellite's Visible Imaging System (VIS) and Earth Camera [*Gjerloev and Hoffman*, 2014]. Care was taken such that the onset brightening developed continuously into a substorm to eliminate pseudo onsets [*Gjerloev and Hoffman*, 2014]. The substorm peak is also identified using Polar VIS images. The peak is a qualitative estimate of the combined intensity of the event and the westward and poleward expansion of the poleward auroral boundary [*Gjerloev and Hoffman*, 2014]. A quiet day and a steady magnetic convection event were also investigated. The quiet day is defined as a day in which no substorms have occurred. The SMC event chosen occurs on 10 February 2008. This event was chosen due to a good distribution of magnetometer stations in the nightside at the start of the event.

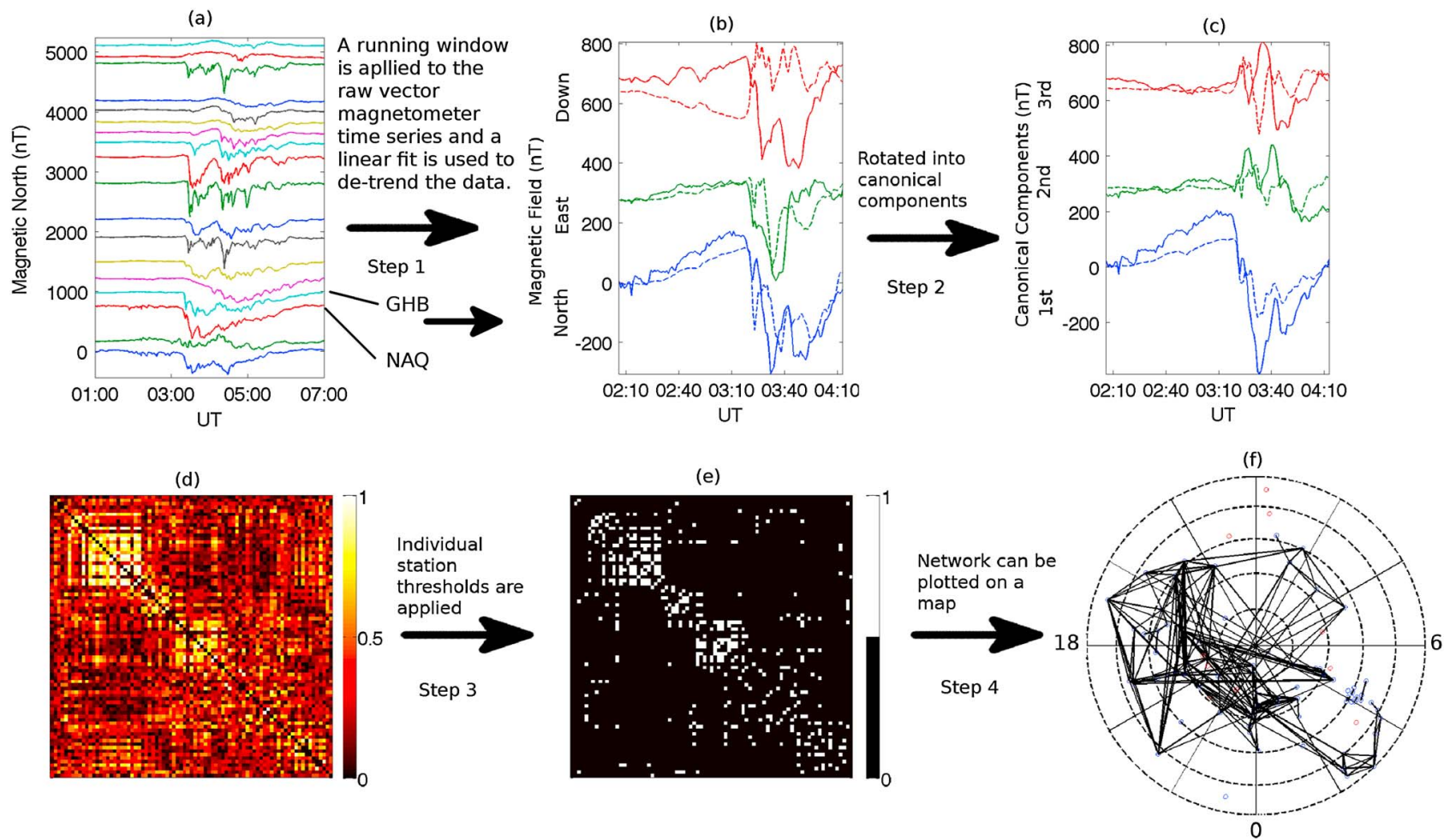

**Figure 1.** An illustration of the process of identifying connections in the network; these steps are outlined in section 2.1. (a) Stack plot of magnetic north time series for stations centred around magnetic midnight during the substorm. The time series are ordered by magnetic latitude. (b) A comparison of a 128 min segment of the north (blue), east (green), and down(z) (red) component of the magnetic field for two stations: GHB (dashed) and NAQ (solid). A linear fit has been used to detrend the data within the window. (c) Canonical correlation is used to form new rotated components that maximize the correlation in the first canonical component for this time window. The rotation is unique for each station pair and time window. Correlation between different canonical components is zero as both the cross and autocovariance for the canonical components matrices are diagonal. (d) The canonical correlation process is repeated for all station pairs, and a correlation matrix, $C_{ij}$, can be formed. The matrix contains the correlation coefficients for the first canonical component. (e) Station-dependent thresholds are applied to $C_{ij}$ to form the adjacency matrix $A_{ij}$. The white squares indicate a connected station pair. Connections can be visualized on a MLAT-MLT map. (f) The magnetic north down view of the Northern Hemisphere, the blue circles indicate active stations, and the red circles are stations for which there are no data at this time. The dashed lines are contours in MLAT, at 50°, 58°, 66°, 74°, and 82°.

### 2.1. Constructing the Network

For each of these events, we form dynamical networks of connected stations in magnetic local time (MLT)-MLAT space in the Northern Hemisphere. To establish whether a pair of stations is connected in the network, we need to identify the time intervals where the correlation between signals from two stations is above a threshold. Connected stations then reflect a shared response to some spatially extended magnetic activity. Whether or not a pair of stations is connected varies with time as the geomagnetic conditions change. The network is formed in the following steps which are illustrated in Figure 1:

*Step 1.* A running 128 min window is applied to the raw vector time series, and the data are detrended in each window with a linear fit. The window is chosen to exceed the timescale in which changes are occurring within the substorm. Linear trends on the window timescale are removed so that the effect of the window on the resultant cross correlation is small. The same running window is used to obtain the time-varying cross correlation (step 2). We use a 128 min window for two reasons: first, a window of sufficient length is needed to have confidence in correlation (results using a shorter window length are included in SI Text S4). Second, 2 h represents the typical global evolution timescale for a substorm. Fluctuations on timescales shorter than the window then can contribute to the amplitude of the cross correlation. Changes in the value of cross correlation on timescales shorter than the window are not resolved. The spatiotemporal scales of the observations fundamentally limit what can be deduced from any measure of cross correlation. Observations are at 1 min resolution, so with a 128 point cross correlation, an equatorial station has moved by ∼3300 km. For a station near the auroral oval (∼70° latitude), the station will have moved ∼1100 km. This roughly determines the spatial resolution implied by the time window.

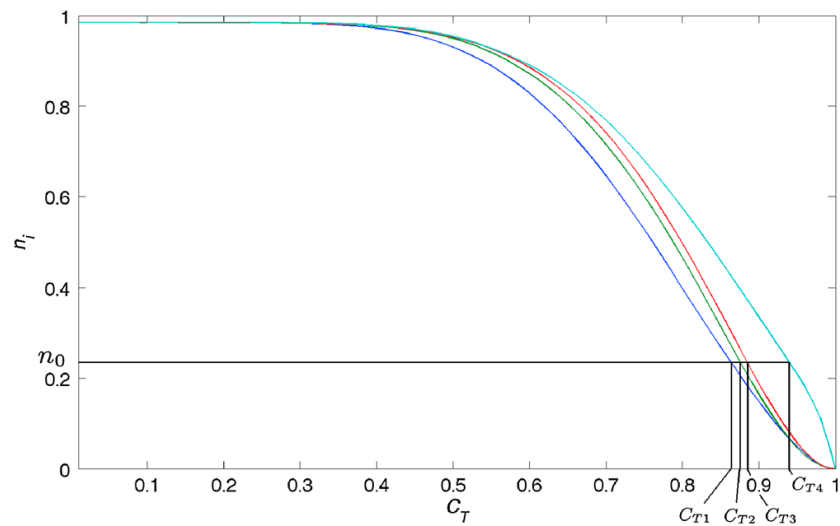

**Figure 2.** The monthly averaged normalized degree for representative stations is plotted as a function of the global threshold, $C_T$. A given normalized degree for the network, $n_0$, gives a corresponding threshold for each station, $C_{Ti}$.

*Step 2.* We use canonical correlation [*Brillinger*, 1975] between the windowed segments of pairs of vector magnetometer time series to establish similarity between pairs of stations as a function of time. We focus on near instantaneous correlation, that is, correlation at zero lag (an example of a 4 min lag network is presented in SI Text S3). We calculate the canonical correlation between the $i$th and $j$th station for all possible station pairs to form a cross-correlation matrix (or weighted network), $C_{ij}(t)$. $C_{ij}(t)$ contains the correlation coefficient for the first canonical component for each station pair.

*Step 3.* Next we threshold $C_{ij}(t)$ to obtain the adjacency matrix $A_{ij}(t)$, which is zero (no connection) or one (connection) for a given pair of stations $ij$. The threshold $C_{Tij}$ in principle has a different value for each pair of stations in the network. We will construct a standardized adjacency matrix such that all stations, on a long timescale (here, 1 month), have the same average degree (or likelihood to be connected to the network). On average each station will, by construction, then be connected to the same fraction of the network as any other station. We define a normalized degree, $n_i(t)$, which is the number of connections a station $i$ has divided by $N(t)-1$, where $N(t)$ is the number of active stations at time $t$. Figure 2 plots the month averaged degree $n_i(C_T)$ obtained by applying a single global threshold $C_T$ across all stations. We plot how $n_i(C_T)$ varies with $C_T$. We standardize our adjacency matrix by finding the threshold $C_{Ti}$ for each station that gives the same fixed degree $n_0$ (when averaged over 1 month). The station-dependent thresholds, used to obtain the time-dependent adjacency matrix, are then $C_{Tij} = \min[C_{Ti}, C_{Tj}]$ (This procedure is described in detail in SI Text S1.).

*Step 4.* Once a network is constructed for each time window we can calculate time-dependent dimensionless parameters that describe the spatial distribution and extent of the correlated behavior. These can then be used to characterize the system.

### 2.2. Network Parameters

Once the threshold for each station pair, $C_{Tij}$, is determined as above, we can obtain the adjacency matrix:

$$A_{ij}(t) = \Phi\left[|C_{ij}(t)| - C_{Tij}\right] \quad (1)$$

so that $A_{ij} = 1$ if station $i$ is connected to station $j$ and is zero otherwise and $\Phi$ is the Heaviside step function. All diagonal elements (self connections) of $A_{ij}$ are set to zero, and in an undirected network the matrix is symmetric, as in the case for correlation calculated at zero lag. Once the dynamical network has been formed, network parameters can be used to quantify its evolution. Given $A_{ij}(t)$, time-dependent global network parameters can be determined as follows:

1. *The normalized total number of connections*,

$$\alpha(t) = \frac{\sum_{i \neq j}^{N(t)} \sum_{j \neq i}^{N(t)} A_{ij}}{N(t)^2 - N(t)}, \quad (2)$$

where $N(t)^2 - N(t)$ is the total number of possible connections in the network. $N(t)$ is the number of active stations taking data, which varies with time.

2. *The average geodesic connection distance* (physical distance), $\delta$, in the network. Note that the average geodesic connection distance here is not the shortest path (or graph geodesic) between two stations [*Newman*, 2010]. A distance separation matrix, $d_{ij}$, is formed that is the geodesic distance separations between all stations. The average connection distance is then

$$\delta(t) = \frac{\sum_{i\neq j}^{N(t)} \sum_{j\neq i}^{N(t)} A_{ij} d_{ij}/(N(t)^2 - N(t))}{\sum_{i\neq j}^{N(t)} \sum_{j\neq i}^{N(t)} d_{ij}}, \quad (3)$$

which is normalized to the average connection distance if all stations were connected. Note, $\delta$ can be $> 1$.

3. $\Theta_{kp}$ *is the number of connections within, and between, two fixed latitudinal bands.* The lower latitude band extends from a lower bound, $L_l = 50°$ MLAT (no station data below this latitude were used), to an upper bound $U_l$. $U_l$ is defined as the upper edge of the auroral oval before onset of the substorm of interest at magnetic midnight. The position of the auroral oval is obtained via visual inspection of Polar VIS images, and $U_l$ is different for each event. The upper latitude band extends from the upper edge of the auroral oval, $L_u = U_l$, to $U_u = 90°$ MLAT. If the latitudinal position of station $i$ is $\theta_i$, then

$$\Theta_{kp}(t) = \frac{\sum_{i\neq j}^{N(t)} \sum_{j\neq i}^{N(t)} A_{ij} \Phi[\theta_i - L_k]\Phi[U_k - \theta_i]\Phi[\theta_j - L_p]\Phi[U_p - \theta_j]}{\sum_{i\neq j}^{N(t)} \sum_{j\neq i}^{N(t)} \Phi[\theta_i - L_k]\Phi[U_k - \theta_i]\Phi[\theta_j - L_p]\Phi[U_p - \theta_j]}, \quad (4)$$

where the subscripts $k$ and $p$ take all values of the indices for the lower and upper bands $u$ and $l$. $\Theta_{uu}$ is then the normalized number of connections between stations in the upper band, $\Theta_{ul}$ the normalized number of connections between the stations in the upper band and stations in the lower band, and $\Theta_{ll}$ the normalized number of connections between stations in the lower band.

### 2.3. Establishing Statistical Significance

We now quantify the likelihood that a connection between a station pair could occur by chance (a "false positive"). For colored noise, canonical correlation is known to produce increasingly high correlation coefficients in the first component for increasing $\beta$, where the power spectrum of the noise inputs varies as $f^{-\beta}$ [*Jackel et al.*, 2001]. To quantify the likelihood of false positives, we construct ten noise surrogate data sets as follows: For each noise surrogate data set, each time windowed segment of the signal for all stations is Fourier transformed. The phases are then randomized, leaving the power spectrum amplitude unchanged, and the resulting signal is then inverse transformed. The same process for forming the network that we apply to the observations is then applied to the surrogate data to obtain an estimate for the number of false positives:

$$F(t) = \frac{\sum_i^{N(t)} \sum_j^{N(t)} f_{ij}}{N(t)^2 - N(t)}, \quad (5)$$

where $f_{ij}$ is the surrogate network and $F$ the normalized total number of connections in the network (i.e., false connections). Ten of these surrogate networks, $f_{ij}$, are formed, and the normalized total number of connections (summed over the surrogate network), $F$, for each surrogate network is found. The average of these 10 surrogate values of the (normalized) total number of false connections can then be averaged to give an estimate of the network false positive number.

## 3. Results

We present results for dynamical networks calculated for four substorms over a 10–12 h interval centered on the substorm onset. We also obtain the networks for a steady magnetic convection event and a "quiet" day (defined by a lack of substorms occurring). The dynamical networks and their parameters are obtained for a 128 min running window with a 126 min overlap; i.e., a new network is calculated every 2 min. The results here focus on canonical correlation networks at zero lag; therefore, the network parameters represent the near-simultaneous response to correlated magnetic activity. A normalized degree for the network $n_0 = 0.05$ is chosen in order to reduce the number of false positives (spurious connections) in favor of allowing more false negatives (unidentified real connections). As a consequence the networks are formed from only the strongest connections in the system.

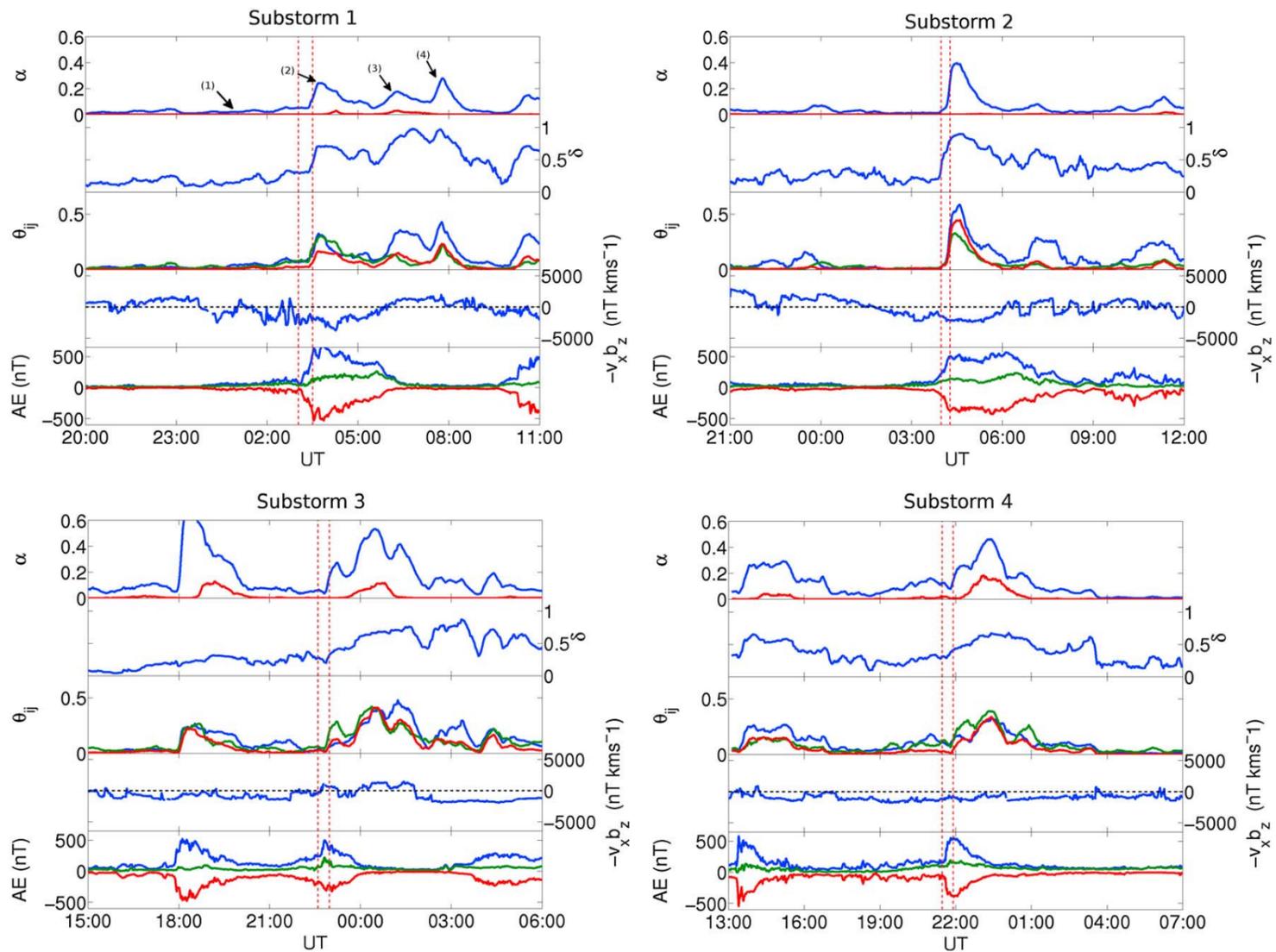

**Figure 3.** The time series of network parameters for four substorms: (top left) substorm 1 03:03 7 January 1997, (top right) substorm 2 04:01 6 November 1997, (bottom left) substorm 3 22:38 3 November 1997, and (bottom right) substorm 4 21:29 16 December 1997. Each subfigure is organized as follows: from top to bottom: (first row) $\alpha$ (blue line), the normalized total number of connections, where the averaged total number of false connections is the red line. (second row) The normalized average connection distance in the network, $\delta$. (third row) $\Theta_{ij}$, the normalized number of connections between MLAT bands $i$ and $j$. There are two MLAT bands, the lower latitude band contains stations between MLAT 50° (no data were used for stations below this point) and the upper edge of the auroral oval (at midnight just before onset) and the upper band between the upper edge and 86° MLAT. The normalized number of connections within the lower band is the blue line, the normalized number of connections within the upper band is the green line, and the normalized number of connections between the lower band and the upper band is the red line. (fourth row) $-v_x b_z$, where $v_x$ is the solar wind velocity along the Earth-Sun line and $b_z$ is the north-south component of the IMF. (fifth row) Also plotted is $AE$ (blue), $AL$ (red), and $AU$ (green). The first vertical dashed red line indicates the onset time, and the second indicates the peak of the substorm. The time of the peak of the substorm is a qualitative estimate based on POLAR satellite images (see section 2). The times (1)–(4) are highlighted in Figure 3 (top left; substorm 1). The networks at these times are plotted in Figure 4. Note, since the occurrence of false positives is independent of geodesic separation length and position, $\delta$, on average, will be 1 and the false connections are evenly distributed in the latitudinal bands.

### 3.1. Network Response to Substorms

Figure 3 plots the time evolution of the network during four substorms. The network parameters are plotted as functions of the time of the leading edge of the correlation window for each realization of the network. Therefore, when comparing with $AE$ and $-v_x b_z$, a range of values equal to the window length must be considered. Both $v_x$ and $b_z$ are propagated solar wind parameters in geocentric solar magnetospheric coordinates, and the data are obtained from the Wind satellite.

Substorms 1 and 2 are isolated events with interplanetary magnetic field (IMF) $b_z$ turning southward 1–2 h before onset. Both substorms have low connectivity in the network before the substorm onset, with any existing connections being short range (i.e., $\delta$ is small). There is a rapid increase in connectivity around onset, accompanied by an increase in $\delta$ above preonset levels for both substorms. Both substorms show an increase in high-latitude connections, low- and cross-latitudinal connections at onset. There is then a gradual decrease in the overall connectivity as the substorms enters the recovery phase. This phase is defined by the slow return of $AL$ to presubstorm levels. For substorm 1, $\delta$ does not decrease during this phase to presubstorm levels and at the end of the substorm there is a resurgence of network activity dominated by low-latitude connections. $\delta$ also reaches its maximum here. We associate this resurgence of activity with the later stages of the recovery phase of the substorm as the correlation window still encompasses a large portion of the recovery phase (the leading edge window time is plotted). Substorm 2 only shows a minor resurgence of activity at the end of the recovery phase consisting almost entirely of low-latitude connections. $\delta$ does only returns to presubstorm conditions after several hours.

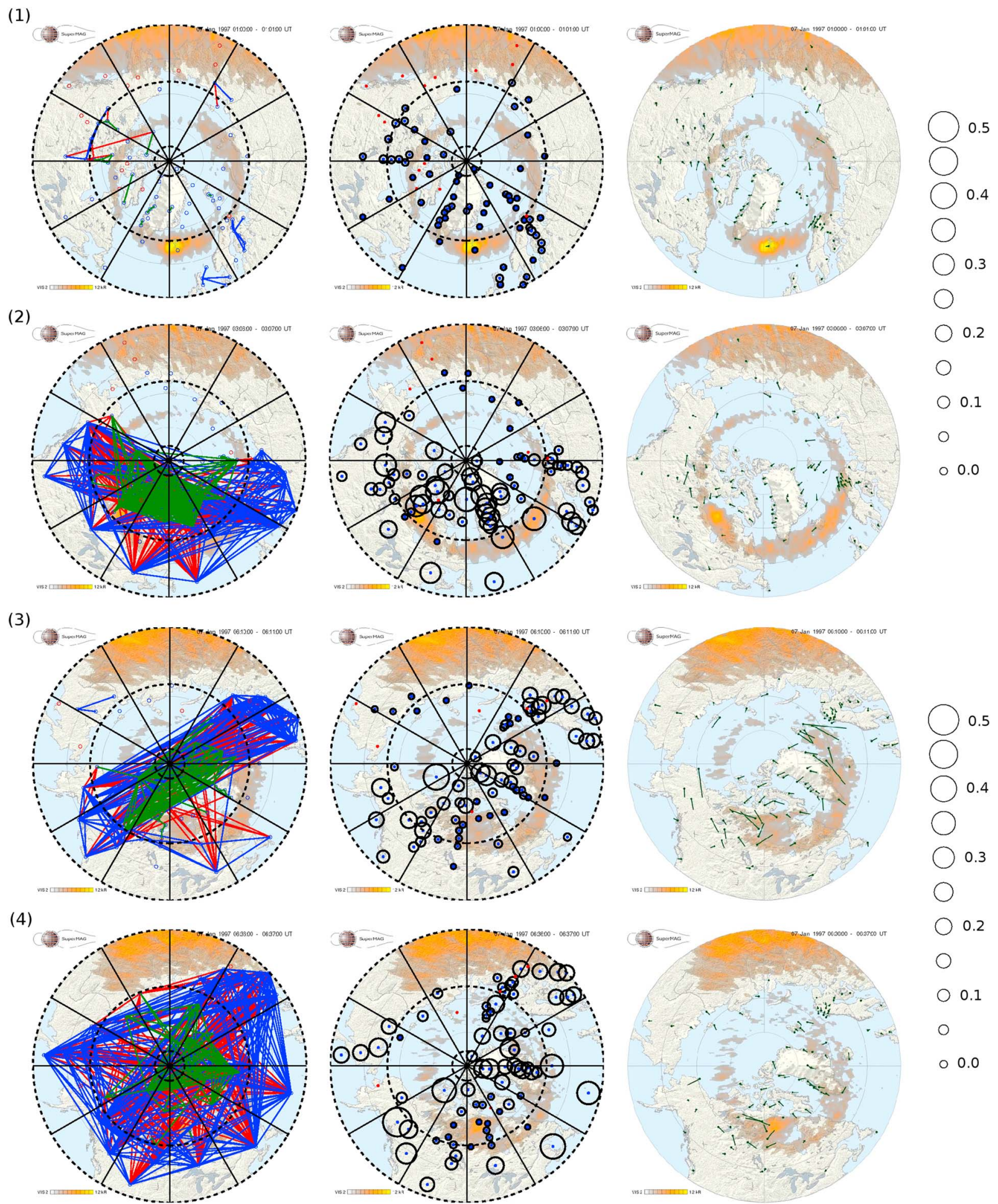

**Figure 4.** (left column) Connection maps for several times during substorm 1, these times are labeled (1)–(4) in Figure 3. The connections are color coded as follows: connections between high-latitude stations only (green) connections between low-latitude stations only (blue) and connections between high- and low-latitude stations (red). The red circles are stations that are not active at that time. (middle column) Displays normalized degree maps for the same times; the radius of the black circles denotes the normalized degree for each given station. (right column) Polar VIS data are also plotted for the same times. The station locations and VIS data, like the network parameters, correspond to the time at the leading edge of the correlation window. The exception to this is at time (4) where no VIS existed for the leading edge time so data from the central window time was used. The black dashed lines correspond to contours in MLAT. The outer most contour corresponds to 50° MLAT, and the next highest corresponds to the boundary between low- and high-latitude stations, 68°. Magnetic midnight is at the bottom of each plot

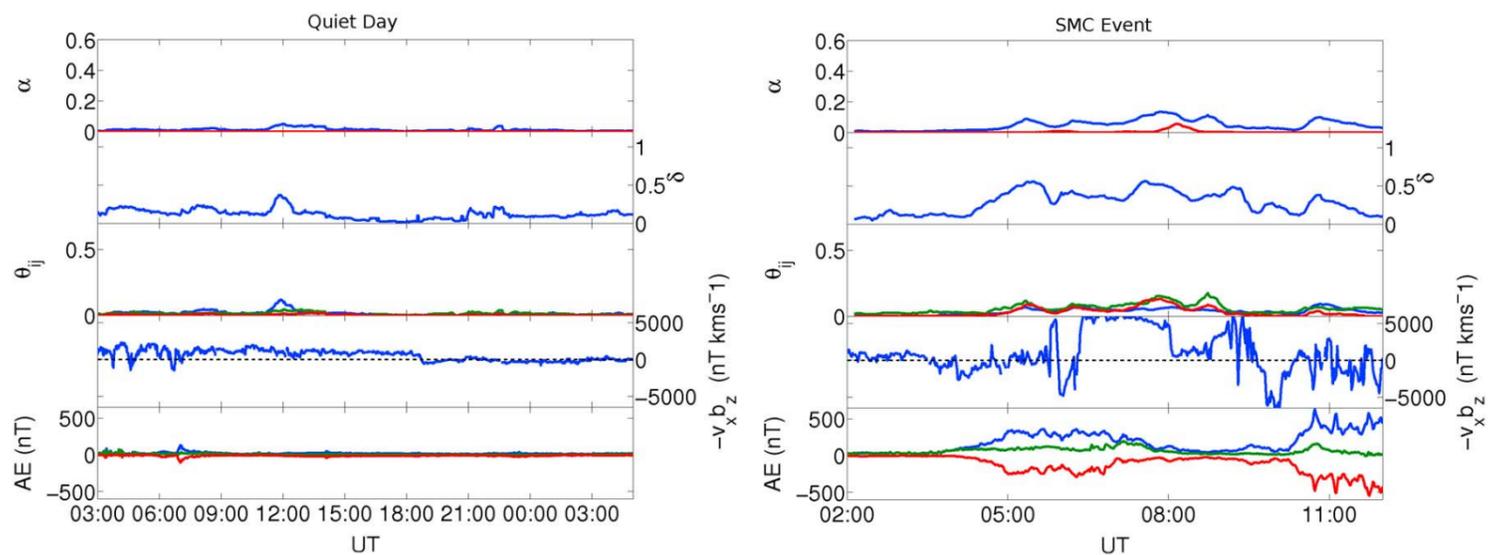

**Figure 5.** The time series of network parameters plotted in the same format as Figure 3 for a (left) quiet day (2–3 February 1998) and a (right) SMC event (05:00–07:00 10 February 2008). The axes scales are the same as in Figure 3.

Substorm 3, unlike the previous two substorms, has a substorm occurring 5 h before the substorm of interest at 18:00 UT. Both substorms can be seen on the plot. IMF $b_z$ remains negative following the end of the previous substorm. There is a strong network response to onset of the substorm, with $\alpha$ reaching 0.28 during the onset peak. At onset there is again a large number of high-latitude connections indicated by $\Theta_{uu}$. There is a second peak in connectivity at the end of the recovery phase where alpha reaches 0.55, i.e., half of all available connections are present. These low-latitude connections dominate here. $\delta$ also reaches maximum here after a slow ramp up during the substorm.

For substorm 4, $b_z$ is southward well before the onset and *AE* is also in a perturbed state. There is a gradual increase in connectivity before the substorm onset with $\alpha$ reaching 0.22 here. The first peak around onset is dominated by high-latitude connections. $\alpha$ continues to increase reaching a maximum of 0.4 during the recovery phase. This phase shows an increase in low-latitude connections. $\delta$ also reaches maximum during the recovery phase after a slow ramp up during the substorm. The likelihood of false connections during this substorms is much higher than the other three substorms (for consistency the same normalized degree, $n_0$, for the network was chosen for all events). The onset peak, however, is largely free of false connections. Substorm 4 shows some activity in the network well before the substorm onset, although it is difficult to identify whether it is associated with this specific substorm or is indicative of unrelated activity.

The magnitude of the response in network parameters, for the test case substorms appears to be largely independent of the magnitude of peak *AE* during the substorms. This seems to indicate that the network is not simply tracking the magnitude of ongoing activity. Sustained *AE* > 300 nT is not necessarily associated with near-simultaneous magnetic activity. This can be seen in substorms 1 and 2 where there is a clear dropoff in connectivity in the network postpeak while AE remains high. This dropoff in $\alpha$ cannot be seen as clearly in substorms 3 and 4 possibly due to the short recovery phase in comparison to substorms 1 and 2.

The networks can be visually represented; we show this for substorm 1 in Figure 4. Figure 4 shows snapshots of the connection maps (left column), maps of the spatial normalized degree distribution (middle column), and Polar VIS data (right column) for times that are indicated in Figure 3 (top left). The connection maps show that the 2 h before onset (1) there is little connectivity in the network, and any existing connections are local. At the onset phase (2) the connection structure is composed of highly concentrated connections at high latitude (the green connections) as well as significant cross-latitudinal connections (the red connections). The connections at this stage are situated around the onset brightening seen in the Polar VIS images. This is seen clearly in the degree maps, stations in the evening sector at high latitudes having large normalized degree. During the recovery phase (3) the correlated behavior shifts to cross connectivity between regions centered around 20 MLT and 8 MLT. At the end of the recovery phase (4) the network is at its most globally distributed, with significant connectivity between the dayside and the nightside. In general, stations with the highest normalized degree are found outside of the expanded auroral bulge region during the recovery phase substorm. However, one cannot directly compare a single snapshot of the auroral visible light emission to the network connection structure, which aggregates information from a 2 h time window.

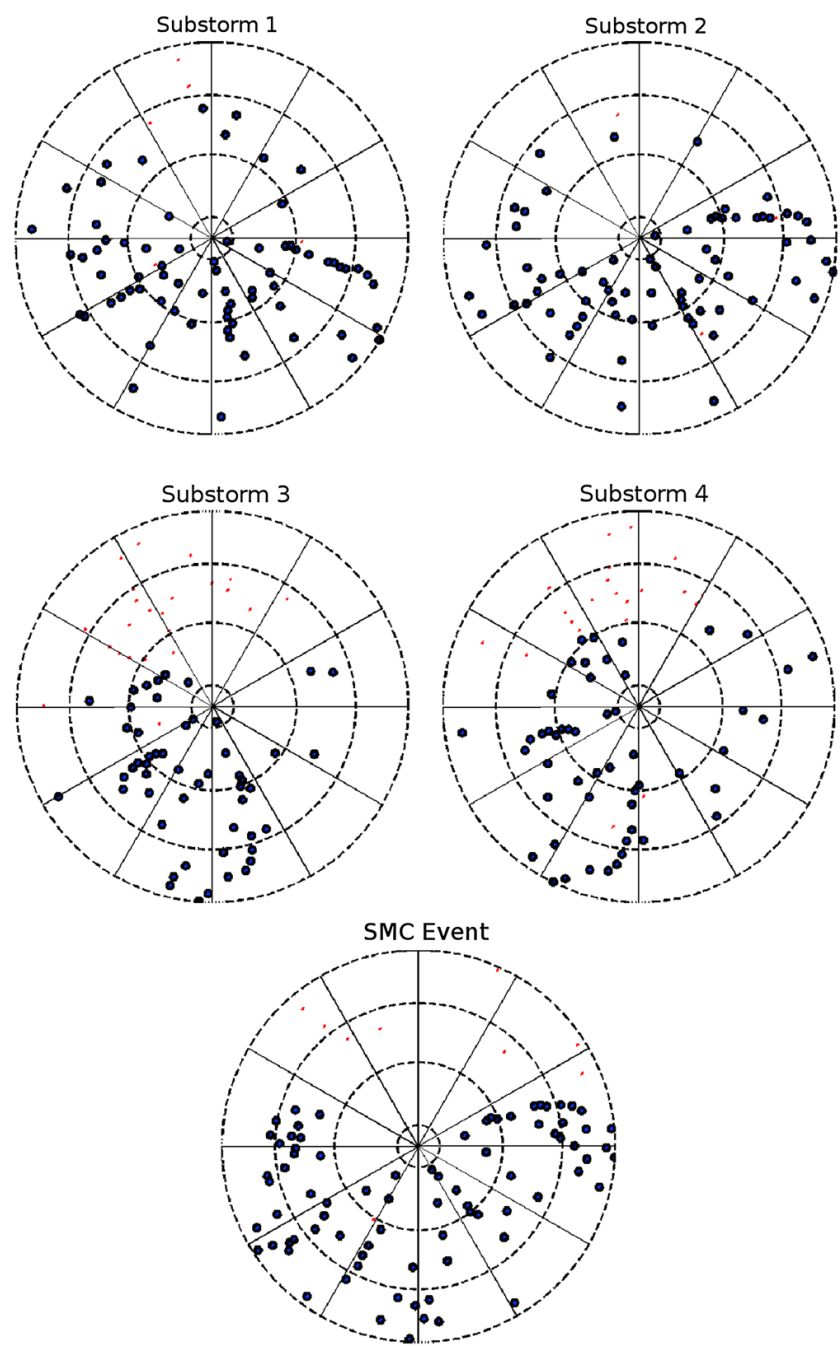

**Figure 6.** Station positions in MLT-MLAT coordinates at the onset of substorms 1–4 and the SMC event. Magnetic midnight is at the bottom of each plot.

### 3.2. Network Response to Other Phenomena

To test the robustness of this approach, we now apply the same methodology to both a "quiet day," defined here by a lack of substorms occurring and a steady magnetic convection event. The results are plotted in Figure 5 where the axes have the same scales and format as Figure 3. The quiet day was selected at random, and there were no constraints on the solar wind conditions. We can see that for the quiet day (Figure 5, left) $\alpha$ does not exceed 0.04, which is significantly lower than that occurring during substorms, and similarly, $\delta$ does not exceed 0.35. There is little discernible network response to the brief step-like southward turnings of the IMF and the subsequent responses in $AE$.

For the SMC event, Figure 5 (right), there is a gradual increase in $\alpha$ at the onset of the event. The increase in connectivity coincides with the increase in $AE$ from ambient levels. $\alpha$ continues to increase and reaches a maximum of 0.11 at the end of the event, which is a factor of 3 less than seen during substorms. Throughout the event connections are dominated by cross-latitude and high-latitude connections. $\delta$ is raised from typical background levels from the onset of the perturbations in $AE$ and reaches a maximum of 0.55. During this event the nightside sector was well represented by the station configuration, from Figure 6 we can see that there is a comparable station configuration, and thus spatial sampling, to substorms 1 and 2.

To summarize, the typical signatures of isolated substorm activity are then as follows:

1. Few connections in the network before onset of substorm.
2. The network exhibits a clear rapid response at onset indicated by an increase in connectivity, $\alpha > 0.22$. High-latitude connections are a key feature of the onset peak; however, low- and cross-latitudinal connections are also present.
3. There is a switch from a high-latitude-dominated connection structure to a low-latitude-dominated connection structures during the later stages of the recovery phase. During the recovery phase $\delta$ usually reaches maximum.
4. The maximum $\alpha$ and $\delta$ reached during the substorms is $\geq 0.32$ and $\geq 0.7$, respectively. In comparison, the maximum $\alpha$ and $\delta$ during the SMC event was 0.11 and 0.55, respectively. Similarly, for the quiet day, the maximum $\alpha = 0.04$ and $\delta = 0.35$.

5. Postsubstorm the network parameters return to their presubstorm state, given that there is no subsequent event shortly after the substorm.

Note that the network parameter values depend on the choice of normalized degree, here $n_0 = 0.05$ (on average 5% of stations are connected). A consistent feature seen in all substorms is the progression from high-latitude connection structures to a low-latitude-dominated connection structure as the substorm enters the recovery phase. From plots of the connection maps for substorm 1 (Figure 4), the high-latitude connections are colocated with the onset brightening. Therefore, the first peak can be associated with the poleward leap and intensification of the auroral electrojet. The recovery phase of the substorm is traditionally associated with the relaxation of the perturbed system back to its ground state [*McPherron et al.*, 1973]. The movement of the closed field line structures in the magnetosphere (associated with latitudes below the auroral oval) during the relaxation to ground state will produce associated currents in the magnetosphere. There is also a significant number of connections to stations in dayside sector during the end of the recovery phase. We interpret the large number of connections during the recovery phase as a clear indication that the substorm electrojet system is coherent on a global scale. If we apply the two-component electrojet concept [e.g., *Kamide and Kokubun*, 1996] this could be interpreted as the convection electrojet system is dominant and the substorm current wedge has ceased to play any significant role.

## 4. Conclusions

In this paper we have outlined a new methodology that uses networks to quantify substorm dynamics. Our results show that substorms can be characterized in terms of the spatial extent and level of cross correlation seen between ground magnetometer stations. Our identification of a consistent network response at onset, which is distinct from events that can appear similar in *AE* (such as SMC events), opens possibilities of using dynamical network analysis as a tool to assist in the identification of substorms using only ground station magnetometers. Information about the spatial distribution of correlation could also be useful in characterizing the dynamics of a substorm, albeit with restricted time resolution due to the correlation window. We have found that there can be a large network response even when the amplitude of the station responses are small but above the noise. Thus, network parameters and geomagnetic indices are complementary, not directly comparable. This paper is a "proof of principle" in that it only explored four test cases for the technique; a more extensive statistical study is needed to fully establish the possibility for network analysis as a method to routinely categorize substorms based on a statistical network. This will be a subject of a future paper. The main results of this study are summarized here:

1. The network exhibits a clear rapid response at onset indicated by an increase in connectivity and average connection distance.
2. There is a strong increase in the number of high-latitude connections at onset. These usually, but not always, dominate the network.
3. Visual inspection of the connection structure at onset shows that they spatially coincide with the location of the onset brightening.
4. The network response to the quiet day and SMC events give quantitatively distinct behavior in the network parameters as compared to the substorm events.

In this study the dynamical network analysis was applied to substorm phenomena. However, with the appropriate window size and lags, the technique, in principle, may be applied to characterize other phenomena that occur in the magnetosphere. Due to the sparseness of the stations in some regions, however, the type of phenomena that can be reliably characterized must be large in its spatial extent.

**Acknowledgments**
This work was supported by the UK Science and Technology Facilities Council, STFC project ST/K502418/1. S.C.C. and J.D. thank Max Plank Institute for Physics of Complex Systems (MPIPKS) Dresden for productive visits. The SuperMAG baselined ground magnetometer station data were obtained freely from http://supermag.jhuapl.edu/.

Michael Balikhin thanks the reviewers for their assistance in evaluating this paper.

## Erratum

In the originally published version of this article, five Supporting Information items were mislabeled and their captions contained errors. These problems have since been corrected, and this version may be considered the authoritative version of record.